\documentclass[twocolumn,prl,superscriptaddress]{revtex4}
\usepackage{graphicx}
\usepackage[final]{epsfig}
\usepackage{ifpdf}
\usepackage{amssymb,latexsym}
\usepackage[mathscr]{eucal}
\usepackage{url}
\usepackage{hyperref}
\usepackage{dcolumn}

\newcounter{mnotecount}[section]

\begin{document}
\hyphenpenalty=800

\title{Fractal Threshold Behavior in Vacuum Gravitational Collapse}

\author{Sebastian J. Szybka}
\affiliation{Astronomical Observatory and Centre for Astrophysics,
  Jagellonian University, Krak\'ow, Poland}
\author{Tadeusz Chmaj}
\affiliation{H.\ Niewodniczanski Institute of Nuclear
   Physics, Polish Academy of Sciences,  Krak\'ow, Poland}
   \affiliation{\small{Cracow University of Technology, Krak\'ow,
    Poland}}

\begin{abstract}
We present the numerical evidence for fractal threshold behavior in the
five dimensional vacuum Einstein equations satisfying the cohomogeneity-two
triaxial Bianchi type-IX ansatz. In other words, we show that a flip of the
wings of a butterfly may influence the process of the black hole formation.
\end{abstract}

\maketitle

\emph{Introduction.}---Critical phenomena in gravitational collapse are an interdisciplinary
field that was initiated by the remarkable work of Choptuik \cite{PhysRevLett.70.9}. It concerns
the study of the basin boundary between two generic states --- space-times with and without black
holes. This bistable behavior shares many properties with the phase transition in statistical
mechanics. It is a well-known fact that, in general, basin boundaries can be either smooth or
fractal, and there is no reason to believe that gravity is special in this context. However,
in more than one hundred papers that were devoted to critical phenomena in gravitational
collapse so far \cite{Gundlach:2002sx,LR2007}, the basin boundary between dispersion and collapse to a black hole is
always smooth, and there is no indication of chaos \cite{footnote1}.
The single counterexample
\cite{Szybka:2003uz} is restricted to the solution with many unstable modes and as such it does
not give a chance to observe fractal threshold behavior directly in the initial value problem.

The theory of chaos in infinite dimensional systems is not fully developed yet, but this field is
of great interest and growing rapidly. In this Letter we provide an example of chaos in reduced
Einstein equations that constitute a system of partial differential equations. We clarify the
hypothesis of Bizo\'n {\it et al}.\ \cite{Bizon:2006qi} and present fractal threshold behavior
that can be interpreted as a chaotic scattering in a critical surface---the surface in the
phase space of initial data that separate collapse to a black hole from dispersion. This surface
is smooth, but it contains three copies of the critical solution, and the basin sets of these
copies have fractal boundaries. In this sense, the chaotic behavior presented here is
double critical and can be directly seen in the dynamical evolution. In our Letter
we estimate
the fractal dimension of the basin boundaries. The fractal dimension is a diffeomorphism
invariant indicator of chaos in general relativity \cite{mixmaster}.

\emph{Setting.}---We consider vacuum gravitational collapse in a very simple setting.
Namely, in order to evade Birkhoff's theorem and save radial symmetry, we take five dimensional
vacuum Einstein equations and reduce the number of degrees of freedom by the BCS ansatz
\cite{Bizon:2005cp}, \cite{Bizon:2006qi}
\begin{eqnarray}
\nonumber ds^2&\!=\!& - A e^{-2\delta} dt^2 + A^{-1} dr^2 \\
&&+\frac{1}{4} r^2 \left[ e^{2B}
\label{metric} \sigma_1^2\!+\!e^{2C} \sigma_2^2\! +\!e^{-2(B+C)}\sigma_3^2\right],
\end{eqnarray}
where $A$, $\delta$, $B$, and $C$ are functions of time $t$ and
radius $r$. One-forms $\sigma_k$ are standard left-invariant one-forms on
$SU(2)$,
\begin{equation}\label{sigma}
    \sigma_1+i\,\sigma_2=e^{i\psi} (\cos{\theta} \;d\phi + i\,
    d\theta), \quad \sigma_3=d\psi - \sin{\theta}\; d\phi,
\end{equation}
where  $0\leq \theta \leq \pi$, $0\leq \phi \leq 2\pi$, and $0\leq \psi \leq 4\pi$ are Euler
angles. The angular part of the metric (\ref{metric}) corresponds to the spatial part of the
metric in Bianchi IX cosmology. $SU(2)$ is dipheomorphic to $S^3$; therefore $B$ and $C$
functions can be interpreted as squashing parameters of $S^3$. $B=C=0$ correspond to the round
$S^3$ and $B=C\neq 0$ to the deformed $S^3$ with one squashing parameter (the so-called biaxial
case). In this Letter we consider the triaxial case where $B$, $C$ are independent functions---two
dynamical degrees of freedom. The vacuum Einstein equations derived for the metric (\ref{metric})
(see \cite{Bizon:2006qi}) possess a discrete $Z_3$ symmetry. It corresponds to the freedom of
permutations of coefficients of one-forms $\sigma_k$ in the angular part of the metric
(\ref{metric}). These permutations are generated by the following transpositions:
\begin{eqnarray}\label{trans}
   &&T_{12}:(B,C)\rightarrow (C,B),\quad T_{23}:(B,C)\rightarrow (B,
   -B-C),\nonumber\\
   &&T_{13}:(B,C)\rightarrow (-B-C, C)\,,
\end{eqnarray}
where the transposition $T_{ij}$  swaps the coefficients of
$\sigma_i^2$ and $\sigma_j^2$ in (\ref{metric}).
Biaxial configurations correspond to the fixed points of these
transpositions: $(B,B)$, $(B,-B/2)$, and $(B,-2B)$. All biaxial
solutions exist in three geometrically equivalent copies.

\emph{Critical phenomena.}---The study \cite{Bizon:2005cp} of the critical phenomena in the model
restricted to biaxial symmetry revealed type-II gravitational collapse with the critical
solution being discretely self-similar ($DSS$). It was also shown \cite{Bizon:2006qi} that in 
triaxial symmetry the phenomenology of critical behavior remains the same. The $Z_3$ symmetry
implies that in the triaxial case the critical solution (hereafter denoted as $DSS_1$---because 
it is discretely self-similar and has one unstable mode) takes one of three equivalent
forms. Let us denote them as $X_1^{(1)}$, $X_2^{(1)}$ and $X_3^{(1)}$. Each of these copies acts
as a codimension-one attractor in the phase space of solutions. Hence, the codimension-one
critical surface $\mathcal{M}_{crit}$ contains three basin sets $\mathcal{M}_i$ ($i=1,2,3$). It
was shown by Bizo\'n {\it et al}.\  \cite{Bizon:2006qi} that the basin boundaries of these sets
are given by the stable manifolds of the discretely self-similar codimension-two attractor $DSS_2$.
Moreover, the authors suggested that the basin boundaries could be fractal.

In this Letter we continue the work that was initiated in \cite{Bizon:2006qi} and
investigate the structure of the basin boundaries of codimension-one attractors $X_i^{(1)}$.
In other words, we study numerically the Cauchy
problem for two-parameter families of initial data. We fine-tune one of the parameters
to the critical surface $\mathcal{M}_{crit}$ and the remaining parameter to the basin boundary 
(this means fine-tuning to the codimension-two attractor $DSS_2$).

One of the problems in carrying on such studies numerically comes from the fact that in order to
cancel two growing modes of $DSS_2$ in general two-parameter initial data, one is forced to use
very high numerical precision. It leads, on the software level, to the necessity of
multiprecision packages. These packages slow down the numerical evolution, and the calculations
become computationally infeasible. The possible solution is to find special two-parameter
initial data that are convenient in the study of the basin boundaries in $\mathcal{M}_{crit}$.
We propose the following procedure for finding such initial data.

Let us consider general initial data that are fine-tuned to $DSS_2$. The solution generated by
such initial data approach $DSS_2$ and, since it is always fine-tuned with a finite precision,
repels from $DSS_2$. However, one can freeze the evolution at the moment the solution is ``nearly''
$DSS_2$. Using this snapshot one can generate another two-parameter family of initial data.

In particular, we consider time-symmetric initial data that were studied in \cite{Bizon:2006qi}:
\begin{equation}\label{ic}
    B(0,r)=p\, f(r), \quad C(0,r)=a\,
    B(0,r),
\end{equation}
where $f(r)=100 \,r^2 \exp[-20(r-0.1)^2]$ is the generalized Gaussian
(for details, see \cite{Bizon:2006qi}). In
  order to fine-tune to $DSS_2$ we
set parameters
\begin{equation}
\begin{array}{ll}
a=0.1411036683285,\\
p=0.09524484187150217214296769,
\end{array}
\end{equation}
and evolve the solution up to $\tilde t=1.038$. Next, we introduce two new parameters $\kappa$,
$\iota$ and rescale $\hat C(\tilde t,r)=\kappa C(\tilde t,r)$, $\hat B(\tilde t,r)=\iota
B(\tilde t,r)$. In this way we create a new, two-parameter ($\kappa$, $\iota$) family of initial
data.

\begin{figure}[htbp]
\includegraphics[width=0.48\textwidth]{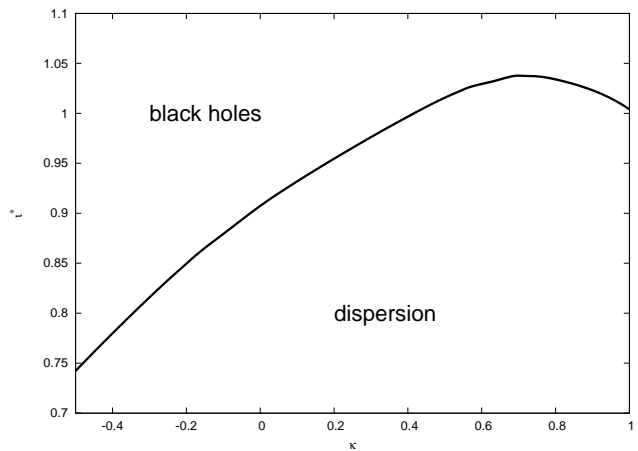}
\caption{The critical surface $\mathcal{M}_{crit}$ presented in the
  phase space of parameters $\kappa$ and $\iota$. All initial conditions
  defined by $\iota^*(\kappa)$ lead to the solutions that approach the
  critical solution $DSS_1$ or attractors of higher
  codimension ($DSS_2$ or possibly others). $\iota\gtrapprox\iota^*$ leads to supercritical solutions
  (with a black hole) and $\iota\lessapprox\iota^*$ gives subcritical solutions (dispersion). 
   The error bars are lower than $10^{-5}$.}\label{fig:cc}
\end{figure}

Hereafter we restrict our analysis only to this particular family.
The critical surface $\mathcal{M}_{crit}$
corresponds to a curve $\iota^*(\kappa)$ in the parameter space. We present
this curve in Fig.\ \ref{fig:cc}.

In order to characterize evolution of initial data given by
$\iota^*(\kappa)$ we define below a map $h:\kappa\in\mathbb{R}\rightarrow
a\in\{1,-\frac{1}{2},-2,0\}$. Generic solutions starting
from $\iota^*(\kappa)$ approach one of the codimension-one attractors
$X_1^{(1)}$, $X_2^{(1)}$, $X_3^{(1)}$ and recover biaxial symmetry $C=aB$ in
one of three equivalent forms: $a=1$, $a=-\frac{1}{2}$, $a=-2$,
respectively. Moreover, there are points that separate basins
of attraction $\mathcal{M}_i$. For a practical convenience we denote them as $a=0$, but we
stress that this is not related to the scaling $C=aB$. The set of all such
points forms the basin boundary $Q$. It was shown in \cite{Bizon:2006qi} that
solutions defined by $Q$ approach the $DSS_2$ solution or possibly 
higher codimension attractors.
The map $h:\kappa\rightarrow a$ describes asymptotic
behavior of solutions, but in finite precision numerical calculations it describes
intermediate dynamics. Such calculations show that if $\kappa_A<\kappa_B$
and $h(\kappa_A)\neq h(\kappa_B)$, and $h(\kappa_A)h(\kappa_B)\neq 0$,
then there exists $\kappa_Q$ such that
$\kappa_A<\kappa_Q<\kappa_B$ and $h(\kappa_Q)=0$. Again, in practical
calculations we mark by $h(\kappa)=0$ all points with an undetermined basin set. For
these points the double precision of the parameter $\iota^*$ is not sufficient
to cancel by a bisection the $\mathcal{M}_{crit}$-transversal
growing mode of $DSS_2$. This mode dominates over the 
$\mathcal{M}_{crit}$-tangential growing mode of $DSS_2$ and the $DSS_1$ behavior is not observed.
In contrast to this, for $h(\kappa)\neq 0$  the $\mathcal{M}_{crit}$-tangential growing
mode of $DSS_2$ takes over and ``pushes'' solutions along $\mathcal{M}_{crit}$.
Such solutions approach one of the copies of $DSS_1$. Finally, since the bisection
has a finite precision, they are also repelled from $\mathcal{M}_{crit}$ along
the unstable mode of $DSS_1$. This behavior was described in detail in \cite{Bizon:2006qi}.

To study the geometry of $\mathcal{M}_{crit}$, especially the geometry of basin sets
$\mathcal{M}_i$, we determined the map $h$ in the numerical experiment. We have collected two
samples corresponding to two scales in the basin boundary. {\it Sample $1$} contains $15001$
points and corresponds to $\kappa\in S_1=[-0.5,1]$. {\it Sample $2$} contains $3201$ points and
corresponds to $\kappa\in S_2=[0.07,0.078]$. All test points were uniformly distributed in
respective intervals.

\begin{figure}[tbp]
\includegraphics[width=0.48\textwidth]{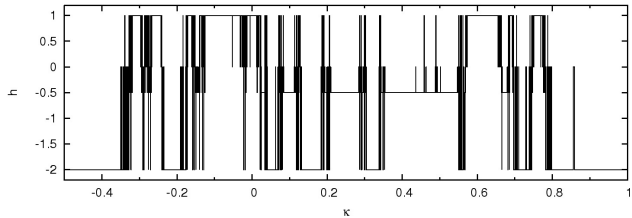}
\caption{{\it Sample $1$}.\ The type of the $DSS_1$ solution as a function of $\kappa$ (plotted as a histogram). $h(\kappa)=1$, $h(\kappa)=-1/2$, $h(\kappa)=-2$
  correspond to $X_1^{(1)}$, $X_2^{(1)}$, $X_3^{(1)}$, respectively, and
  $h(\kappa)=0$ corresponds to points with an undetermined basin.}\label{fig:frac}
\end{figure}

\begin{figure}[bp]
\begin{center}
\includegraphics[width=0.48\textwidth]{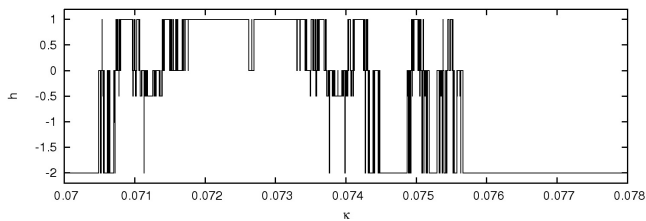}
\end{center}
\caption{{\it Sample $2$}.\ Part of Fig.\ \ref{fig:frac} magnified. The complex structure is
  seen at a different scale.}\label{fig:mfrac}
\end{figure}

It is clearly seen in Fig.\ \ref{fig:frac}
that the basin boundaries have a complex structure \cite{footnote2}.
Magnifying a part of the boundary reveals a new structure
presented in Fig.\ \ref{fig:mfrac}.
Therefore, one should
expect that an intersection of the basin boundary $Q$ with tested sets $S_1$,
$S_2$ has nontrivial fractal dimension $0<dim(S_i\cap Q)<1$, where $i=1,2$. In order to estimate the
fractal dimension we follow  \cite{1983PhLA...99..415G,fractalbb} and calculate the uncertainty
dimension \cite{footnote3} (from now on $dim$ denotes the uncertainty
dimension).

Let $S$ be a one-dimensional set in one-dimensional parameter phase space (we
have one free parameter $\kappa$). The probability
that any two random points $\kappa_A$, $\kappa_B$ separated by a distance $\epsilon$ belong to different
basins $h(\kappa_A)\neq h(\kappa_B)$ scales as $P(\epsilon)\sim \epsilon^{1-dim(S\cap Q)}$. Testing many
pairs of points for several values of $\epsilon$ one can fit $P(\epsilon)$ to
the power law and estimate $dim(S\cap Q)$. The wider the range of $\epsilon$ we
consider (assuming $\epsilon$ is small enough), the
higher precision we obtain.

\begin{figure}[tp]
\includegraphics[width=0.48\textwidth]{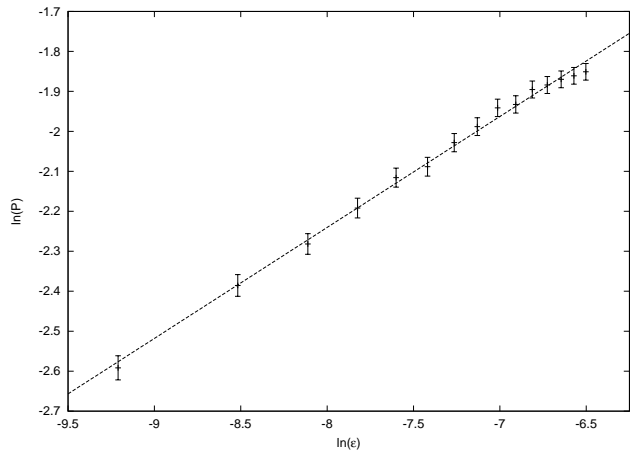}
\caption{{\it Sample $1$}.\ $P$ is a probability that any two random points $\kappa_A$, $\kappa_B$ separated by a distance $\epsilon$ belong to different
basins $h(\kappa_A)\neq h(\kappa_B)$. The uncertainty dimension was estimated
to be $dim(S_1\cap Q)=1-w=0.722\pm 0.006$, where $w$
is a slope of the linear fit above.}\label{fig:4}
\end{figure}

\begin{figure}[bp]
\includegraphics[width=0.48\textwidth]{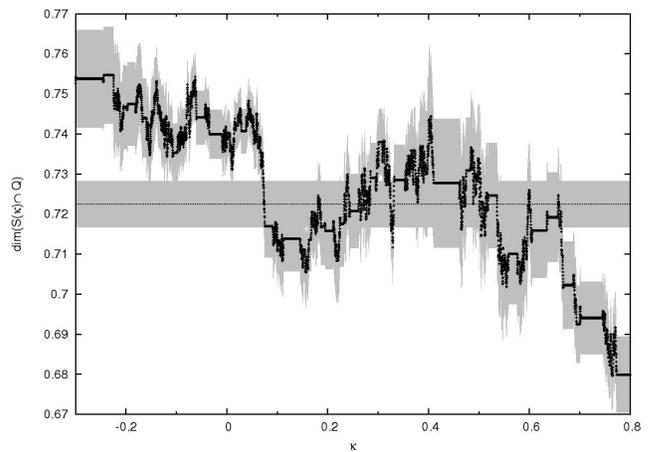}
\caption{
The uncertainty dimension $dim(S(\kappa)\cap Q)$ varies with $\kappa$. The
  set $S(\kappa)=[\kappa-0.2,\kappa+0.2]$ contains $4001$ points for each
  value of $\kappa$. The horizontal dotted line indicates averaged
  uncertainty dimension $dim(S_1\cap Q)$. The gray area represents errors.}
\label{fig:5}
\end{figure}

\begin{figure}[tbp]
\includegraphics[width=0.48\textwidth]{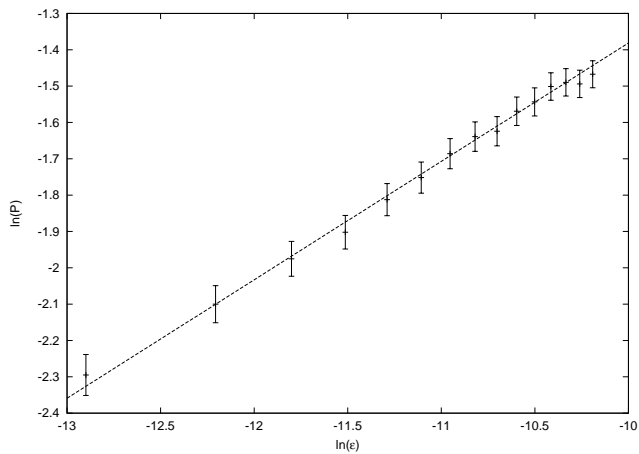}
\caption{{\it Sample $2$}.\
The uncertainty dimension was estimated to be
$dim(S_2\cap Q)=1-w=0.680\pm 0.006$, where $w$
is a slope of the linear fit above.}\label{fig:6}
\end{figure}

The fit to the power law for {\it Sample $1$} presented in Fig.\ \ref{fig:4} reveals 
that $dim(S_1\cap Q)=0.722\pm 0.006$, therefore the
basin boundary is fractal. However, the quantity estimated above is an
``averaged'' fractal dimension. In order to see this, let us consider a set
$S(\kappa)=[\kappa-0.2,\kappa+0.2]$ and $dim(S(\kappa)\cap Q)$ for $\kappa$
such that $S(\kappa)\subset S_1$.
It follows from Fig.\ \ref{fig:5} that the uncertainty dimension $dim(S(\kappa)\cap Q)$ varies with
$\kappa$, so the investigated structure seems to be a multifractal rather
than a monofractal. Such hypothesis is supported by the fractal dimension of
{\it Sample $2$}, namely $dim(S_2\cap Q)=0.680\pm 0.006$ (see the linear fit
in Fig.\ \ref{fig:6}).
It follows from the definition of the uncertainty
dimension that $S_2\subset S_1\Rightarrow dim(S_1\cap Q)\geq dim(S_2\cap
Q)$. The estimated values satisfy this inequality as expected.
The smaller $\epsilon$ we take, the more precise bisection we need and more quickly the
numerical double precision of $\iota^*$ is exhausted. Therefore, {\it Sample $2$} contains ca.\
$18\%$ of points with an undetermined basin in contrast to only ca.\ $7\%$ in case of {\it Sample
$1$}. Clearly, the finite precision of the bisection makes the number of points with an undetermined
basin scale dependent and may have a greater effect on the fractal dimension of {\it Sample $2$}. The
error bars shown in Figs.\ \ref{fig:4} and \ref{fig:6} indicate statistical errors
\cite{1983PhLA...99..415G}.

The multifractal objects can be more completely characterized by the singularity spectrum, but
such analysis of our system would involve more precise data and an enormous computational
power. Let us mention that the computer resources needed to collect {\it Sample $1$} and {\it
Sample $2$} involved over six dozens of processors with the integration time measured in weeks.

The direct analogy to finite dimensional dynamical systems suggests that the
$DSS_2$ solution plays a role of a
strange repeller \cite{bgo}, but we do not have sufficient dynamical evidence to support such claim. To
the best of our knowledge, the nature of
solutions that drive the dynamics in a fractal basin boundary remains unknown
in the context of  partial differential equations
\cite{tc}. A construction of the $DSS_2$ solution in \cite{Bizon:2006qi} is a first
example that may give some insight into
this problem.

\emph{Conclusions.}---In summary, we have presented the numerical results that indicate the presence of 
fractal basin boundaries in the critical surface in five dimensional vacuum gravitational
collapse. Moreover, the data support the hypothesis that basin boundaries are multifractals.
We note that all three copies of the critical solution ($X_1^{(1)}$,
$X_2^{(1)}$ and $X_3^{(1)}$) are geometrically equivalent. Therefore, one
should not expect to see chaos in the black hole mass scaling of supercritical
solutions. The solutions behave chaotically in the nearly double-critical
regime, and then settle down to the Minkowski space-time or the black
hole. This final state is
not sensitive to initial conditions, but the way in which it is
approached is, indeed, chaotic.

\emph{Acknowledgments.}---We thank Piotr Bizo\'n for a
helpful discussion and Zbis{\l}aw
Tabor for letting us adapt his code. The numerical experiments presented here
were performed on machines of Interdisciplinary Centre for Mathematical and
Computational Modelling at Warsaw University (ICM, Grant No.\ G32-6),  Academic
Computer Centre Cyfronet AGH (Grant No.\ MNiSW/IBM\_BC\_HS21/UJ/104/2007), Astronomical Observatory UJ, Institute of Physics
UJ and Henryk Niewodnicza\'nski Institute of Nuclear Physics PAN.
This research was also supported by MNiSW Grant No.\ \hbox{1 P03B 012 29}. \vskip -0.7cm
\vspace{0.5cm}

\bibliographystyle{apsrev}

\begin{thebibliography}{9}
\bibitem{footnote2} All three possible types of behavior are present
in the neighborhood of the boundary. It would
be interesting to verify if the basin boundary has the so-called Wada property---any point
which is on the boundary of one basin set is also simultaneously on the
boundary of all other basin sets.
\end{thebibliography}

\end{document}